\documentclass{emulateapj}

\def\sun{\odot}

\slugcomment{to appear in the Astrophysical Journal Letters}

\lefthead{Tomita et al.}
\righthead{Applicability of the multiple regression analysis to the variable component in the near-infrared region for type I AGNs}

\begin{document}

\title{Multiple regression analysis of the variable component in the near-infrared region 
for type 1 AGN MCG+08-11-011}

\author{Hiroyuki Tomita\altaffilmark{1,2},
   Yuzuru Yoshii\altaffilmark{3,6},
   Yukiyasu Kobayashi\altaffilmark{4},
       Takeo Minezaki\altaffilmark{3},
           Keigo Enya\altaffilmark{7},
    Masahiro Suganuma\altaffilmark{4},
         Tsutomu Aoki\altaffilmark{1},
         Shintaro Koshida\altaffilmark{4,5},
and         
       Masahiro Yamauchi\altaffilmark{4,5}
}

\altaffiltext{1}{Kiso Observatory, Institute of Astronomy, School of Science,
                 University of Tokyo, 10762-30
                 Mitake, Kiso-machi, Kiso-gun, Nagano 397-0101, Japan}
\altaffiltext{2}{Email: tomitahr@cc.nao.ac.jp}
\altaffiltext{3}{Institute of Astronomy, School of Science,
                 University of Tokyo, 2-21-1
                 Osawa, Mitaka, Tokyo 181-0015, Japan}
\altaffiltext{4}{National Astronomical Observatory, 2-21-1
                 Osawa, Mitaka, Tokyo 181-8588, Japan}
\altaffiltext{5}{Department of Astronomy, School of Science,
                 University of Tokyo, 7-3-1
                 Hongo, Bunkyo-ku, Tokyo 113-0033, Japan}
\altaffiltext{6}{Research Center for the Early Universe,
                 School of Science, University of Tokyo, 7-3-1
                 Hongo, Bunkyo-ku, Tokyo 113-0033, Japan}
\altaffiltext{7}{Japan Aerospace Exploration Agency, 3-1-1
                 Yoshinodai, Sagamihara, Kanagawa, 229-8510, Japan}


\begin{abstract}

We propose a new method of analysing a variable component for type 1 active galactic nuclei (AGNs) in the near-infrared wavelength region.  This analysis uses a multiple regression technique and divides the variable component into two components originating in the accretion disk at the center of AGNs and from the dust torus that far surrounds the disk. Applying this analysis to the long-term $VHK$ monitoring data of MCG+08-11-011 that were obtained by the MAGNUM project, we found that the $(H-K)$-color temperature of the dust component is $T = 1635$K $\pm20$K, which agrees with the sublimation temperature of dust grains, and that the time delay of $K$ to $H$ variations is $\Delta t\approx 6$ days, which indicates the existence of a radial temperature gradient in the dust torus.  As for the disk component, we found that the power-law spectrum of $f_\nu \propto \nu^\alpha$ in the $V$ to near-infrared $HK$ bands varies with a fixed index of $\alpha\approx -0.1$ --- +0.4, which is broadly consistent with the irradiated standard disk model. The outer part of the disk therefore extends out to a radial distance where the temperature decreases to radiate the light in the near-infrared.

\end{abstract}

\section{Introduction}

The existence of massive black holes at the center of accretion 
disks in AGNs is well accepted in current AGN research.  The so-called big 
blue bump, which is observed in the UV to optical wavelength region 
of spectral energy distribution (SED) of many AGNs, is regarded as an 
evidence of the accretion disk (e.g., Shields 1978; Malkan \& Sargent 1982; 
Czerny \& Elvis 1987).  Based on another distinct feature of near-infrared 
bumps in their SED, the existence of a surrounding dust torus, which is 
heated to the 
sublimation temperature of dust by the accretion disk, is also accepted now 
(e.g., Sanders et al. 1989; Barvainis 1987; Kobayashi et al. 1993).  

In practice, however, it is known that composite SEDs constructed from a 
huge number of AGNs obtained by large surveys show their optical colors 
to be significantly redder than predicted by the standard accretion disk model 
(see Francis et al. 1991 for the composite SED from the low-redshift bright 
quasar survey (LBQS); Vanden Berk et al. 2001 for the composite SED from the 
Sloan digital sky survey (SDSS)).  Since AGNs are superimposed on host 
galaxies, how serious this discrepancy is depends on how reliably the 
nuclear SED has been separated from the SED of the galaxy component. 

In addition to the problem of SED decomposition, a possible contamination 
of near-infrared light from accretion disk and dust torus further complicates 
the interpretation of the SED.  In fact, it is reported that the SED of the accretion 
disk extends to the near-infrared wavelength region (Kishimoto et al. 2005; Minezaki et al. 2006), 
and that the temperature gradient exists in the dust torus of which the inner 
part is hotter than the outer part (e.g. Barvainis 1987; Glass 1992; Alloin 
et al. 1995). These two factors cause similar effects such that variations 
at shorter near-infrared wavelengths are less delayed behind the optical 
variations, when compared to variations at longer near-infrared wavelengths. 
Therefore, unless the above factors are taken into account simultaneously in 
the analysis, the time delay of near-infrared variations behind the optical 
is liable to be underestimated.

In order to overcome such complications as well as the problem of SED 
decomposition, we propose a new method of analysing the variable SEDs of 
accretion disks and dust tori, based on a multiple regression technique.  
In section 2, we describe the assumptions and methods of the analysis.  
In section 3, we 
apply the analysis to the long-term $VHK$ monitoring data of MCG+08-11-011 
obtained by the MAGNUM project (Yoshii 2002; Kobayashi et al. 2003; Yoshii et al. 2003).  
Finally, in section 4, we describe the applicability of the analysis and discuss the results.

\section{Multiple Regression Analysis}

\subsection{Overview of the Analysis}

Let $F_\lambda(t)$ be the observed flux at wavelength $\lambda$ 
and time $t$, then it is decomposed as 
\begin{equation}
F_\lambda(t)=F_\lambda^{var,disk}(t)+F_\lambda^{var,dust}(t)+
F_\lambda^{const}, 
\end{equation}
where the first two are the variable components from the accretion disk 
and dust torus, respectively, and the last component is a constant, partly
 offset including the contribution from the host galaxy. 

Kobayashi et al. (1993) showed that all 14 QSOs in their sample have 
almost the same blackbody spectra of 1500K. This indicates that the flux 
in the near-infrared wavelength region is mainly radiated from the 
innermost part of the dust torus, where the temperature is near the sublimation 
temperature of dust grains.  Hence, as we observe for each AGN, the 
near-infrared light curves in different bands have almost the same shape 
and overlap each other by shifting the lag time between the variations in 
different bands (e.g., Suganuma et al. 2006).  When there is a short
lag time of $\tau$ between variations in two near-infrared bands, 
$H(1.7\mu m)$ and $K(2.2\mu m)$ for example, which are intrinsic to the 
dust torus, the situation considered here corresponds to 
\begin{equation}
F_H^{var,dust}(t-\tau)/F_K^{var,dust}(t)=a .
\end{equation}
Winkler et al. (1992) and Winkler (1997) claimed that the optical colors of 
the variable component for Seyfert galaxies remain constant as their brightness
changes, while recent studies report that QSOs get bluer as they get brighter 
(e.g., Trevese \& Vagnetti 2002; Giveon et al. 1999). For our analysis 
focusing on Seyfert galaxies below, we then assume that the variable 
component synchronized 
with optical variation, which comes from the accretion disk, has constant 
optical $V(0.55\mu m)$ to near-infrared $H$ and $K$ colors, i.e., 
\begin{equation}
F_H^{var,disk}(t)/F_V(t)=b \;\;\; {\rm and} \;\;\; 
F_K^{var,disk}(t)/F_V(t)=c . 
\end{equation}
Substitution of equations 2 and 3 in equation 1 therefore gives
\begin{equation}
F_H(t-\tau)-b\times F_V(t-\tau)=a\times (F_K(t)-c\times F_V(t))+d , 
\end{equation}
where $a$, $b$, $c$, and $d$ are the constant parameters to be determined 
by the fit to the $VHK$ monitoring data, based on a multiple regression 
technique described below.

\subsection{Details of Calculation}

For calculation of the multiple regression, Jefferys (1980) provides the 
most generalized multivariate least-square process. This process allows errors 
to appear in all variables, which is desirable for astronomical observations. 
In practice, we adopted the Mardquardt method described in Jefferys (1981), 
which is robust for the case that the initial trial for the fit is far from 
the true solution.

The calculation of the regression with a given value of lag time $\tau$ 
requires interpolation of the monitoring data in time, because such data 
are usually discrete.  We here adopt the linear interpolation scheme, and 
smaller weights are assigned to the data produced with large interpolation.  
We estimate the uncertainties of the interpolation, using the structure 
function (SF) of AGN variabilities (Collier \& Peterson 2001; Suganuma et 
al. 2004):
\begin{equation}
\sigma_{\rm SF}^2={\rm SF}(l)=\frac{1}{N(l)}\sum_{i<j}[ f(t_i)-f(t_j)]^2-2\sigma^2 ,
\end{equation}
where the summation runs over all $(i,j)$-pairs, $l=t_j-t_i$ is the length of 
interpolation, 
$N(l)$ is the number of pairs at $l$, $f(t)$ is the flux at time $t$, and 
$\sigma$ is the mean observational error for the data used for the SF 
calculation. We adopted a length nearer to the neighboring data point as the length of interpolation $l$. While we naturally use the observational error as the weight 
for the data with no interpolation, we adopt the larger of the SF error or 
the observational error as representing the uncertainty of the interpolated 
data in the regression analysis. 
In order to determine whether or not the fit well represents the observations,
 the $\chi^2$-test is performed. The $\chi^2$-value is computed by 
doubling equation 8 of Jefferys (1980), and this doubled value follows the 
$\chi^2$ distribution of $N=m-4$ degrees of freedom, where $m$ is the number 
of data sets.

Finally, the value of lag time $\tau$ that gives the best fit is estimated 
at the peak of multiple correlation coefficient that measures the strength 
of association between the independent and dependent variables.

\vspace{0.5cm}
\centerline{{\vbox{\epsfxsize=7.5cm\includegraphics[angle=-90,width=7.5cm]{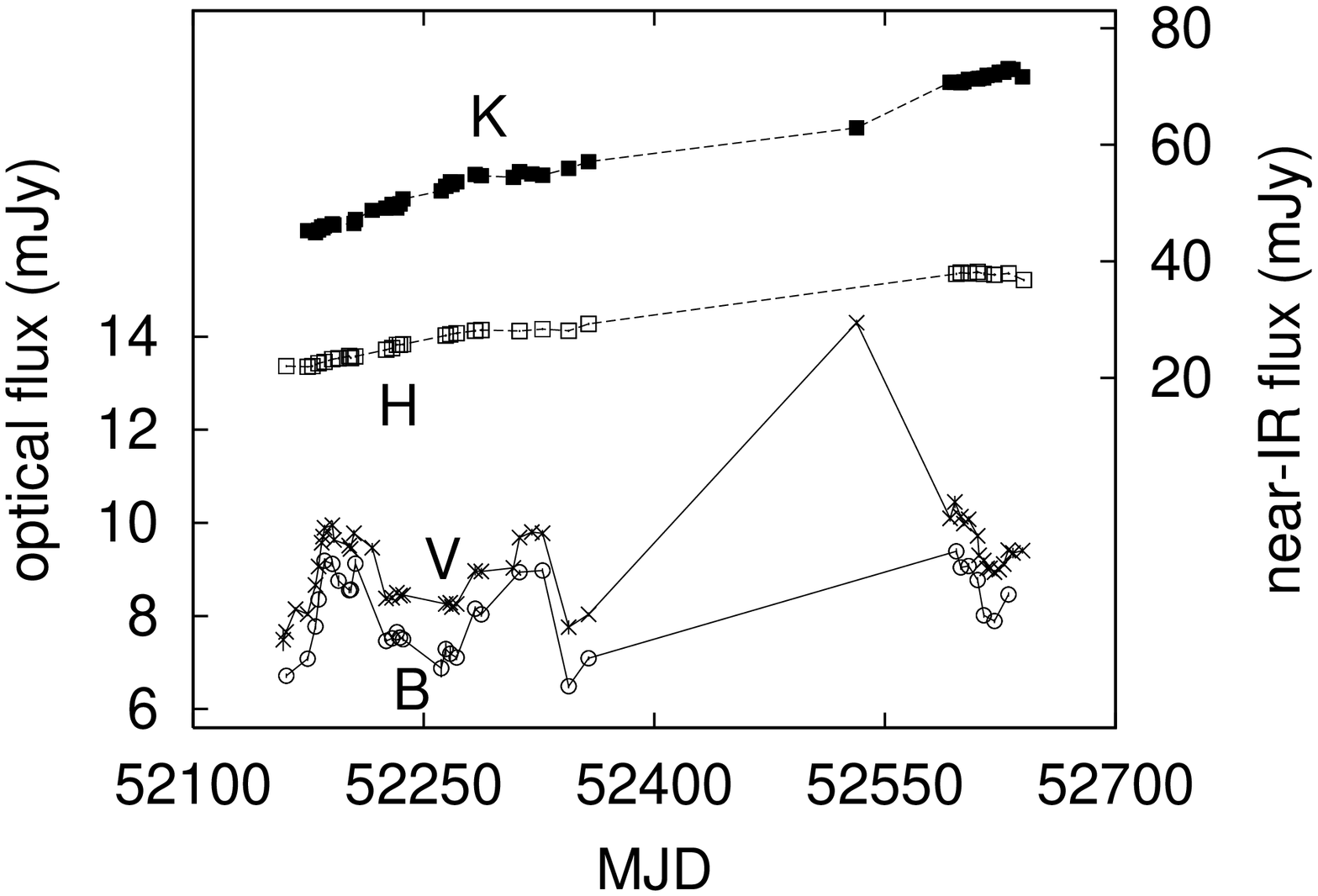}}}}
\figcaption{Light curves of MCG+08-11-011 for the period from 8 September 2001 
to 31 December 2002 (UTC).  Shown are the monitoring data for the $B$ band 
(open circles), $V$ band (crosses), $H$ 
band (open rectangles), and $K$ band (filled rectangles), where the flux from 
the host galaxy has been subtracted.\label{f1}}
\vspace{0.5cm}

\section{Application of the Analysis}

The $BVHK$ monitoring data for MCG+08-11-011 were obtained by the MAGNUM 
project in the period from 8 September 2001 to 31 December 2002 (Figure 1).
Aperture photometry was used throughout. An aperture size of 8.3 arcsec was
adopted for all images. 
The flux of MCG+08-11-011 was measured relative to a reference star 
($\alpha_{2000}= 05^h 54^m47.^s9, \delta_{2000}=+46^{\circ}23'5''$) whose 
flux was absolutely calibrated using Landolt standard stars (Landolt 1992) in 
$BV$ and Hunt standard stars (Hunt et al. 1997) in $HK$  
during the monitoring observations. The host galaxy flux was estimated using  
the GALFIT program (Peng et al. 2002) and was subtracted from the measured 
flux of MCG+08-11-011. The estimated magnitudes of the host galaxy were 
16.5 mag at $B$, 15.5 mag at $V$, 12.0 mag at $H$ and 11.5 mag at $K$. 
A small error of 0.003 mag caused by the seeing effect, arising from 
the profile difference between the AGN and the reference star, was taken 
into account. Galactic 
extinction was corrected according to Schlegel, Finkbeiner, \& Davis (1998).

We performed the regression analysis comparing the observed $H$-band light 
curve with the predicted $H$-band light curve in equation 4. We adopted 
only the data set where observations in $V$, $H$ and $K$ were carried out 
on the same night in order to decrease the effect of interpolation. 
The parameters used for SF$(l)=S_0l^\beta$ were $S_0=0.033\; \mathrm{mJy}^2$ 
and $\beta=1.1$ for $V$, and $S_0=0.0077\; \mathrm{mJy}^2$ and $\beta$=2.0 for 
$K$.  Zero-correlation was assumed between the data in any two bands, and  
the standard error was used for the error of the fitted parameters.

We carried out the regression analysis for a range of $\tau$ from $-10$ to 
60 days. The value of the multiple correlation coefficient was calculated as a 
function of $\tau$, and its peak was found to occur at $\tau=6.0$ days with 
the $\chi^2$-value of 12.5 for $m=26$ (Figure 2). The reduced chi-square 
value becomes $\approx$ 0.5, which indicates that the fit is good enough 
to represent the observations. The regression at 
$\tau=6.0$ days yielded that the $(H-K)$-color temperature for the dust 
component is 1635$\pm$20K, and the index of power-law SED for the accretion 
disk component ($f_\nu \propto \nu^\alpha$) is $\alpha=+0.36_{-0.25}^{+0.34}$ 
from $V$ to $H$ and $\alpha=-0.06_{-0.24}^{+0.35}$ from $V$ to $K$. 

\vspace{0.5cm}
\centerline{{\vbox{\epsfxsize=7.5cm\includegraphics[angle=-90,width=7.5cm]{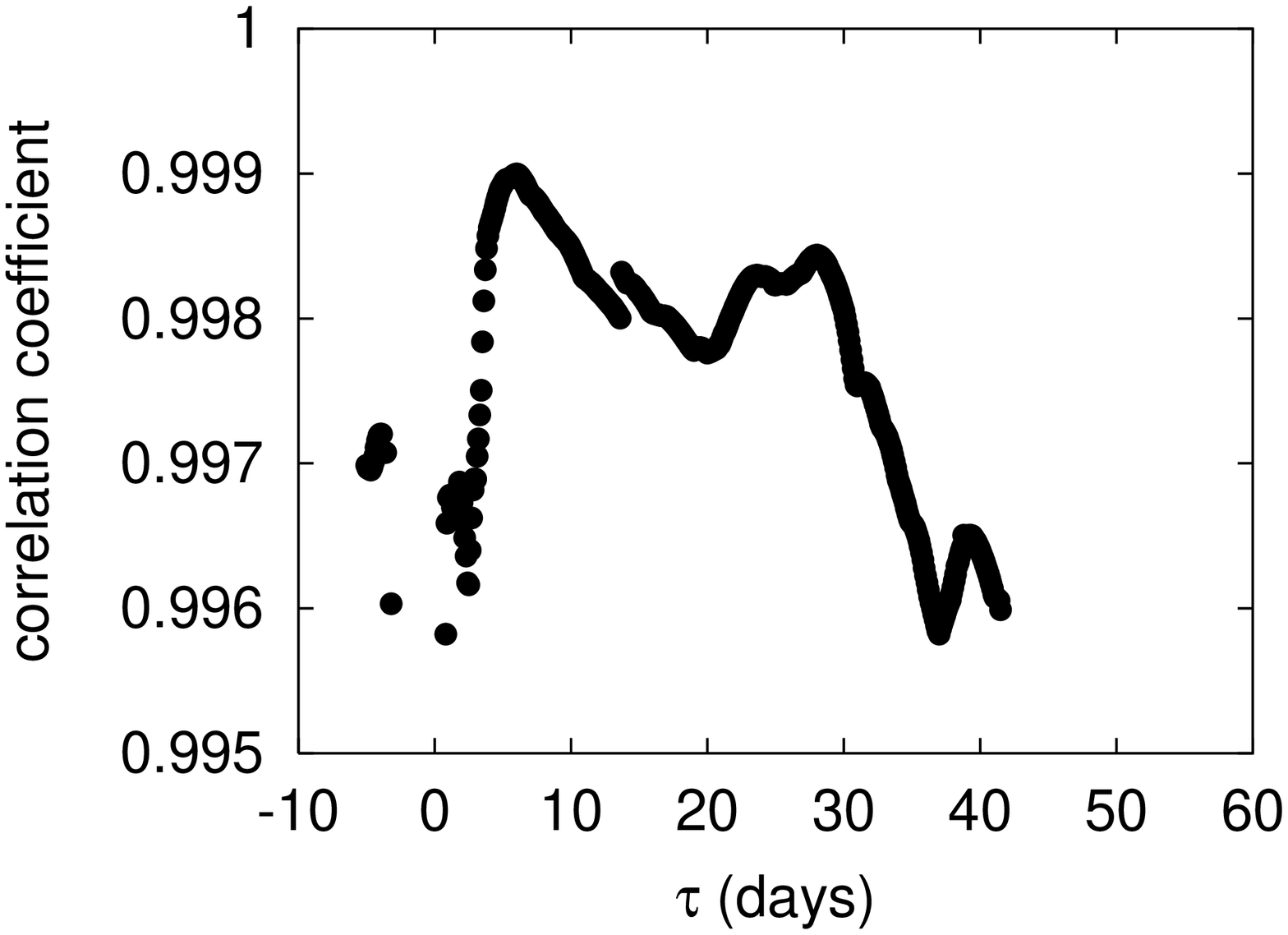}}}}
\figcaption{Multiple correlation coefficient as a function of lag time $\tau$ 
between variations in the $H$ and $K$ bands for the case of MCG+08-11-011.  
The coefficient value is peaked at $\tau=6.0$ days.\label{f2}}
\vspace{0.5cm}

\section{Discussion}

The $(H-K)$-color temperature of 1635K here derived for MCG+08-11-011 agrees 
well with the sublimation temperature of $1500-1800$K for graphite/silicate 
grains (Salpeter 1977; Huffman 1977).  Hence, this component represents the 
innermost part of the dust torus heated to the sublimation temperature by the 
accretion disk.

Following Winkler et al. (1992), we performed the flux variation gradient 
(FVG) analysis for the $BV$ monitoring data (Figure 3). We converted the 
FVG $(B-V)$-color to the index of the power-law SED and obtained 
$\alpha$=+0.21$\pm$0.10.  Combining this and the results in section 3 
from the regression analysis, we found that the power-law SED with 
$\alpha=-0.1\sim +0.4$ prevails in the optical to near-infrared wavelength 
region (Figure 4).  

For comparison, we computed theoretical colors of the 
standard accretion disk model (Shakura \& Sunyaev 1973) with a black hole 
mass of $M=10^7 M_{\odot}$ and an outer disk radius of $R=10^4 R_g$, where 
$R_g = GM/c^2$.  Figure 4 shows $\alpha$ as a function of frequency $\nu$ 
for two cases of accretion rate $\dot{M}=0.1 M_{\odot}$yr$^{-1}$ 
(upper thick line) and $0.001 M_{\odot}$yr$^{-1}$ (lower thick line). 
These predicted SEDs have an approximate value of $\alpha=1/3$, and agree 
with our finding of $\alpha=-0.1\sim +0.4$ for MCG+08-11-011. 

The standard disk model has a critical difficulty for explaining the 
variability.  Although many UV/optical monitoring observations have so 
far been made, all of them have failed to detect significant lags between UV and 
optical variations, except for NGC7469 (Peterson et al. 1998; Wanders et al. 
1997; Collier et al. 1998) and Ark564 (Collier et al. 2001).  The expected lag 
time is too short for the energy to flow at sound speed within the standard 
viscous disk.  However, such a lag time is sufficient for the incident 
radiation to transfer the energy to any distant parts of the disk 
(Krolik et al. 1991).  Hence, the irradiated accretion disk model has been 
invoked, where the central X-ray energy source illuminates a geometrically 
thin disk, and UV/optical light is radiated at the outer part of the disk. 
The predicted spectrum of the 
irradiated standard disk model has an index of $\alpha=1/3$ at the radii 
where the height of the X-ray source is negligible compared to the radial 
distance from the center. This spectrum is exactly the same as that of the 
standard viscous disk model. Furthermore, the detected wavelength-dependent 
lag time agrees with $\tau\propto\lambda^{4/3}$ in the UV to optical region 
for NGC7469 and Ark564, which is the same as the prediction for the irradiated 
standard disk model (Collier et al. 1998; Collier et al. 2001).  Consequently, 
we conclude that the power-law SED obtained by our analysis 
reinforces evidence that irradiated standard disks exist in the center 
of AGNs. 

\vspace{0.5cm}
\centerline{{\vbox{\epsfxsize=7.5cm\includegraphics[angle=-90,width=7.5cm]{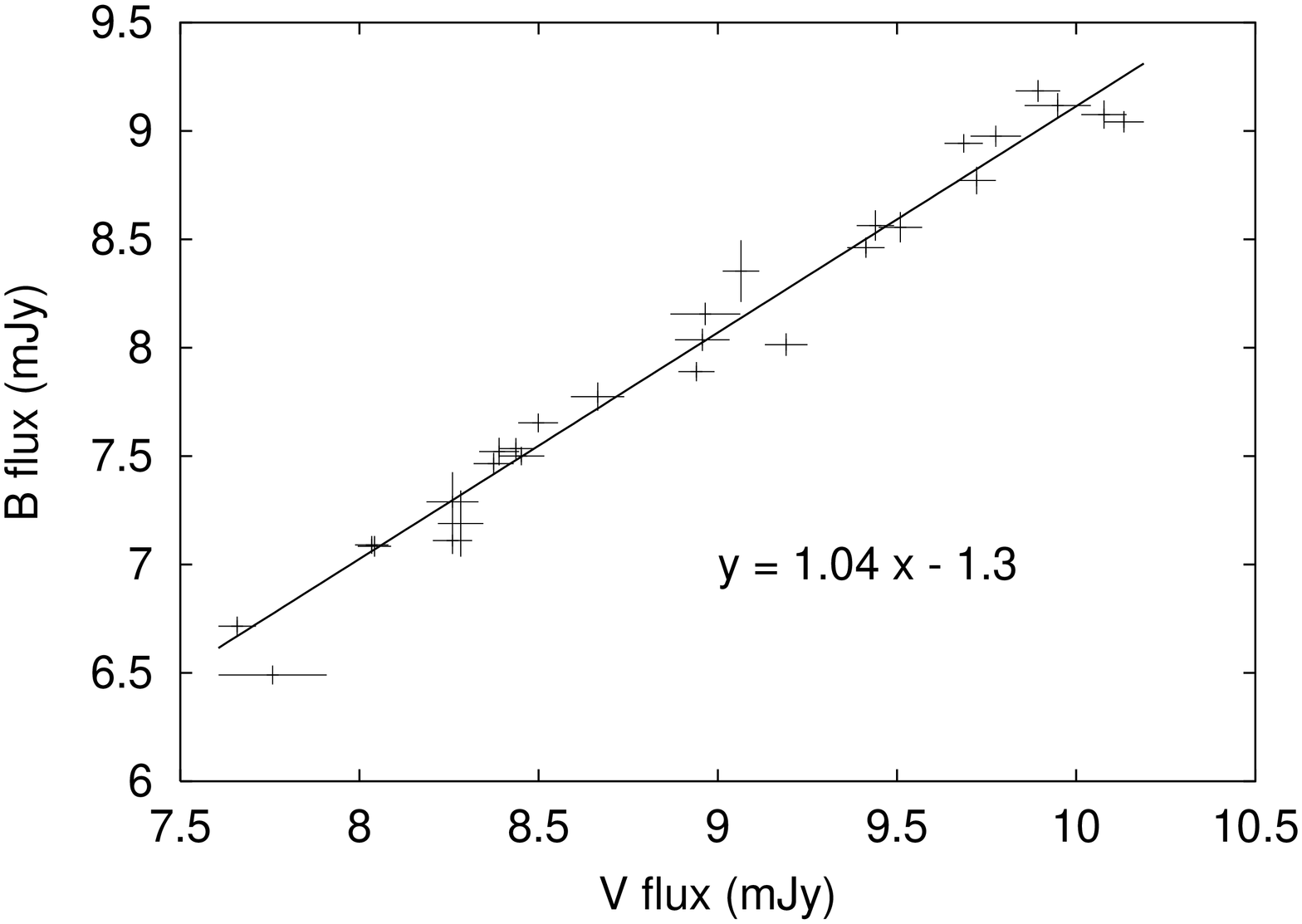}}}}
\figcaption{The $B$-flux to $V$-flux diagram for MCG+08-11-011. 
The straight line is the least-square fit to the data.}

\vspace{0.5cm}
\centerline{{\vbox{\epsfxsize=7.5cm\includegraphics[angle=-90,width=7.5cm]{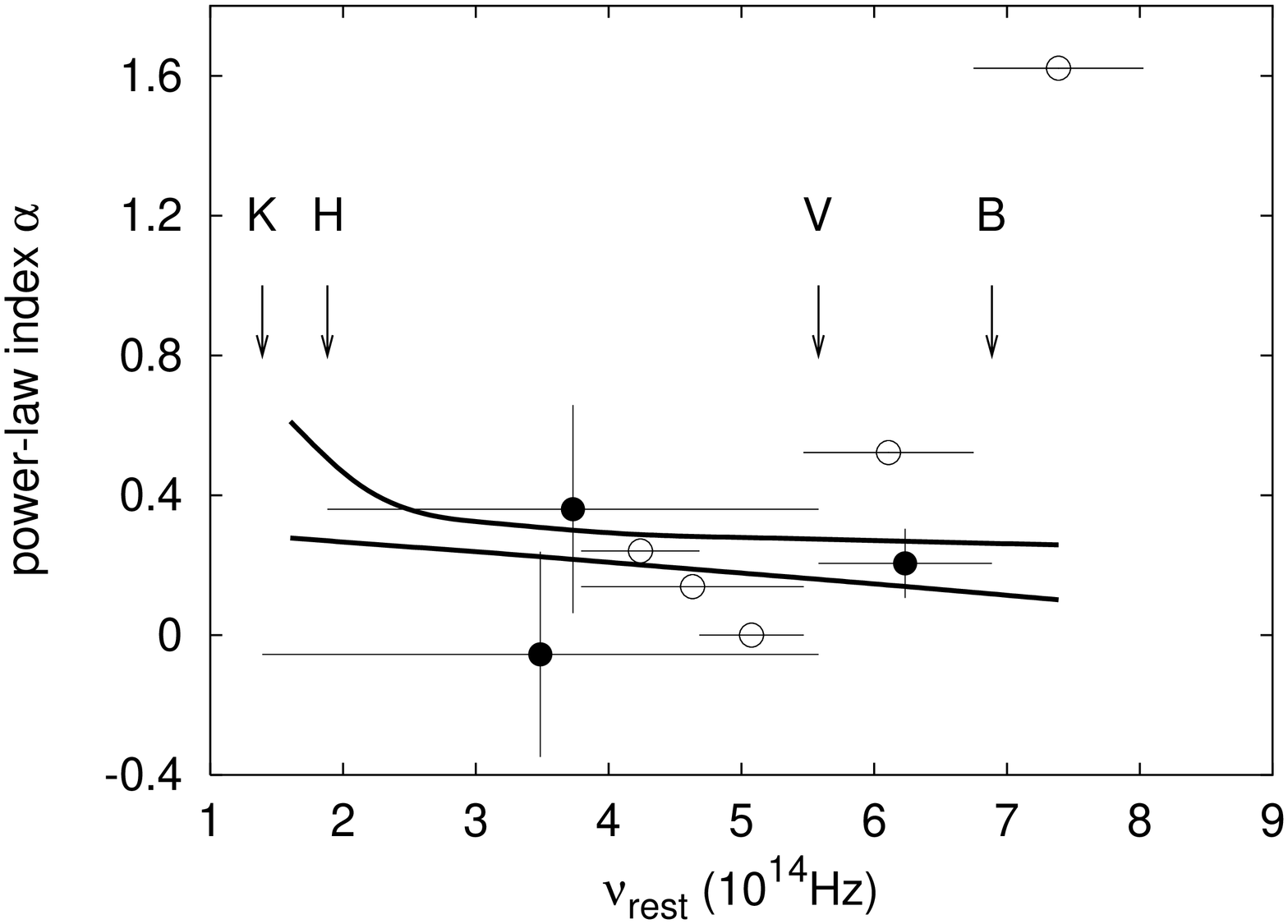}}}}
\figcaption{Values of power index $\alpha$ plotted against the rest-frame 
frequency of radiation. Filled circles are our results for MCG+08-11-011 
from the data of $BV$, $VH$ and $VK$. Vertical bar indicates  the estimated 
error of 
$\alpha$, and horizontal bar corresponds to the wavelength range between 
two bands used for calculation of $\alpha$. For reference, open circles show 
the values of $\alpha$ for other AGNs by Winkler et al. (1992), and two thick 
lines show the predictions of two standard accretion disk models. 
See text for detail.}

\vspace{0.5cm}

Our analysis showed that the power-law SED extends to the near-infrared 
wavelength region. By the standard accretion disk model, the near-infrared 
$K$-emitting region in the disk is located at a radial distance of $R\approx 
10^3R_g$, from the center of typical Seyfert 
galaxies, which is close to the 
inner part of the broad line region (BLR) where high-ionization lines 
originate.  
Collin \& Hure (2001) showed that the disk would become gravitationally 
unstable at $R>R_{crit}\approx 2 \times 10^4R_g(M/10^7M_{\odot})^{-0.46}$. 
Based on this 
estimation, the $K$-emitting region in the disk could be gravitationally 
stable, and the existence of such a stable extended disk could distort the Kepler 
motion of BLR gas clouds.  This may explain the significant scatter of 
observed virial relationships of the broad-emission lines, in accord with the 
suggestion of Peterson et al. (2004) that the possible existence of some 
additional components may affect the BLR gas motion. 
Thus, near-infrared observations of AGNs bring information of physical 
coupling between the extended disk and the BLR.  Particularly, in the case 
of AGNs with high Eddington ratio, the $K$-emitting region in the disk 
would be further away from the centers of AGNs, and the outer disk 
edges might be observable (Collin et al. 2002; Kawaguchi 2003; 
Kawaguchi et al. 2004). 

The accretion disk component contributes to the near-infrared flux by 
15-30\% in the $H$ band and by 15-25\% in the $K$ band (Figure 5).  
Since the observational errors are of order 1\% only, these disk 
contributions are significant, and cannot be ignored in deriving  
accurate and reliable results from near-infrared observations of AGNs. 
For a long time, however, separation of accretion disk components 
in the near-infrared 
region has been considered difficult, because the thermal emission 
from the dust torus component is the dominant source there.  In this context, 
the multiple regression analysis for multicolor monitoring observations is 
indeed a promising method for extracting the accretion disk component from 
near-infrared data, and will be useful for investigating the physical conditions 
and emission mechanisms of both components of accretion disk and dust torus 
simultaneously. 


\vspace{0.5cm}
\centerline{{\vbox{\epsfxsize=7.5cm\includegraphics[angle=-90,width=7.5cm]{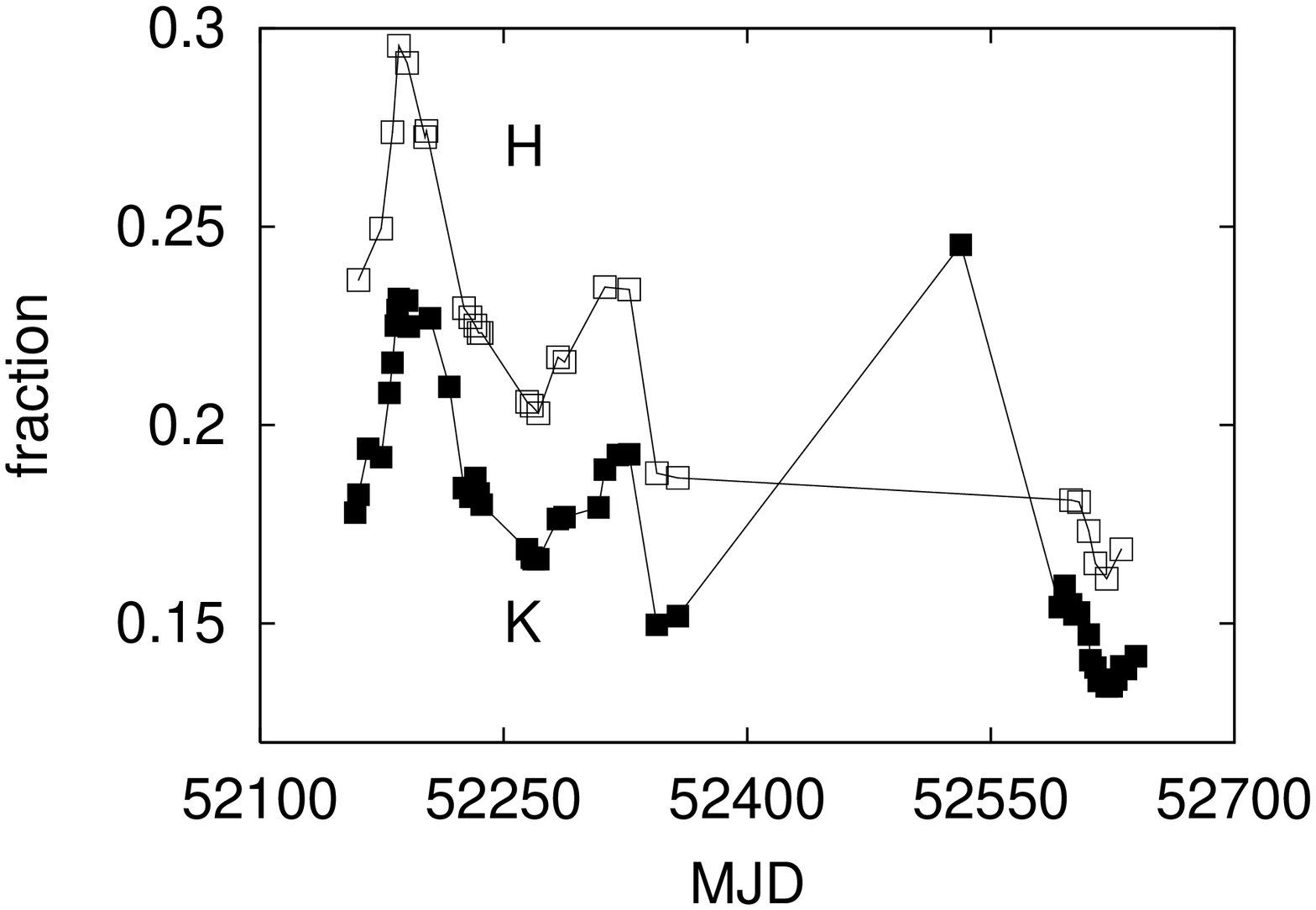}}}}
\figcaption{Fractional contribution of near-infrared flux from the accretion 
disk component throughout the monitoring observations of MCG+08-11-011. Shown 
are the values of $F_\lambda^{var,disk}(t)/F_\lambda(t)$ for the $H$ band 
(open squares) and for the $K$ band (filled squares).}
\vspace{0.5cm}

\acknowledgements
We thank B. A. Peterson for helpful discussions. This research has been supported partly by the Grant-in-Aid of Scientific Research (10041110, 10304014, 11740120, 12640233, 14047206, 14253001, 14540223, 16740106) and COE Research (07CE2002) of the Ministry of Education, Science, Culture and Sports of Japan.



\end{document}